\documentclass{PoS}

\title{Direct photon-charged hadron coincidence measurements in STAR}

\ShortTitle{Direct photon-charged hadron coincidence measurements}

\author{\speaker{Ahmed Hamed for the STAR Collaboration}\\
        for the STAR Collaboration\\
	Texas A\&M University, USA\\
        E-mail: \email{ahamed@tamu.edu}} 

\abstract{The multiplicities of charged particle with transverse momentum (3 $<$ $p_T^{h^{\pm}}$ $<$ 16 $\mathrm GeV/c$) 
and pseudorapidity ($\mid\eta\mid$ $\leq$ 1.0) associated with direct photons and $\pi^{0}$ of high transverse momentum (8 $<$ $p_T^{\gamma, \pi^{0}}$ $<$ 
16 $\mathrm GeV$) at mid-rapidity ($\mid\eta\mid$ $\leq$ 0.9)  
have been measured over a broad range of centralities for $Au+Au$ collisions and $p+p$ collisions 
at $\sqrt{s_{NN}}$ = 200 $\mathrm GeV$ in the STAR experiment. 
A transverse shower-shape analysis in the 
STAR Barrel Electromagnetic Calorimeter Shower Maximum Detector is used to discriminate between the 
shower of single photon and that of photons from the decays of high $p_T$ $\pi^{0}$s. The relative azimuthal positions of 
the associated particles with respect to the trigger particle are constructed, and the associated charged hadron yields per 
direct $\gamma$ are extracted.
An agreement is observed between the measured suppression for 
direct $\gamma$-trigger associated-particle yields in $Au+Au$ compared to that in $p+p$  
and theoretical calculations, although the uncertainties are large.
Within the current uncertainties, the suppression is similar to the previously observed suppression in single-particle 
yields as well as in hadron-triggered associated-particle yields.}

\FullConference{High-pT Physics at LHC -09\\
		 February 4- 4 2009\\
		 Prague, Czech Republic}

\begin{document}
\section{Introduction}

High-$p_{T}$ direct $\gamma$, photons unaccompanied by additional hadrons, produced in high-energy 
collisions are of special importance for the formalism of perturbative Quantum 
Chromodynamics (pQCD). This is because the point-like coupling of the
$\gamma$ to the hard interaction, in principle, makes this process ideally free from uncertainties inherent in jet reconstruction ( as
in the case of jet production) or in fragmentation of partons into hadrons (as in inclusive hadron production), and hence a clean probe of
the hard-scattering dynamics. The direct $\gamma$-hadron azimuthal correlation measurement has been suggested as a powerful tool to 
quantify the energy loss in the medium created at relativistic heavy ion collisions [1]. Unlike quarks and gluons, 
the photon does not have to fragment into hadrons and can be directly observed as a final state particle carrying 
the total energy of the parton.
Furthermore, in heavy-ion collisions these particles are formed early in the collisions 
dynamics (prompt production) and therefore represent an ideal tool for probing the early stage 
of the evolution of a short-lived ($\approx 10 fm/c$) and small size ($\approx 5 fm^{3}$) medium created in the collisions.

The existence, in QCD, of a gluon-photon Compton process, $g+q \rightarrow \gamma +q$, leads to the prediction that the $\gamma /
\pi^{0}$ ratio is large compared to $\alpha$ = 1/137, since the photon is produced directly, whereas the $\pi^{0}$ is a fragment of a
quark or gluon jet coming from the subprocess $q+q \rightarrow q+q$ etc. For single particle spectra quantitative predictions have been
made by several authors [2]. The suppression of $\pi^{0}$ in central $Au+Au$ collisions compared to the $p+p$ scaled by the number 
of nucleon-nucleon scatterings [3] make it possible to perform high-statistics measurements with experiments at RHIC. 
At RHIC energy at mid-pseudorapidity, the dominant process is the q+g scattering for the particles produced 
with $p_{T} \approx 5-15 GeV/c$ [4].  

There are a variety of mechanisms for producing high-$p_{T}$ photons. At the lowest order (LO) in the strong running 
coupling constant processes $O(\alpha \alpha_{s}(Q^{2}))$ the direct photon production is given by the Born-level subprocesses
$q(\bar{q})+g \rightarrow q(\bar{q})+\gamma$ and $q+\bar{q} \rightarrow g+\gamma$ [5,6].  
However, the computation of the next-to-leading order (NLO) contributions yields
$O(\alpha_{s}^{2})$ corrections resulting from the subprocesses $q+q \rightarrow q+q+\gamma$, $q+\bar{q} \rightarrow g+g+\gamma$, 
$q(\bar{q})+g \rightarrow q(\bar{q})+g+\gamma$, and from virtual corrections to the Born-level subprocesses [5,6]. In this regard, a
calculation [7] of the $q+q \rightarrow q+q+\gamma$ subprocesses shows that it provides only a small correction to the basic
$2 \rightarrow 2$ subprocesses given above. It is widely anticipated that at very large values of $p_{T}$ the LO subprocesses 
should dominate.

On the other hand, the contributions due to the Final State Radiation (FSR) , the effects coming from 
the intrinsic constituent motion, and the Initial State Radiation (ISR) could obscure the parton initial energy. 
Although high-$p_{T}$ photons could be produced by a quark fragmentation via
bremsstrahlung (fragmentation photons); but, since the photon takes only a fraction of 
the parton's momentum, this is not the most efficient way to create a high-$p_{T}$ photon. 
Furthermore, the photon will in this case be accompanied by additional hadrons from
the fragmenting quark. This contribution will be suppressed by approximately a factor of $\alpha$ with respect to the other major source of
background, single $\pi^{0}$ rate. Although this suppression will be offset somewhat by the fact that the $q \rightarrow \gamma$ 
fragmentation is flatter than that
for $q \rightarrow \pi^{0}$, the net contribution is still small. 
Finally, due to the parton intrinsic motion $k_{T}$ [8], an  
enhancement in the single-particle spectra is observed [9]. However, for the $\gamma$-jet case,    
the effect of the parton $k_{T}$ is greatly reduced compared to single photon spectra [10,11].                                                           

In the dominant QCD process of Compton-like scattering $(q+g \rightarrow q+\gamma)$, indeed 
the photon transverse momentum balances the parton initial transverse energy $p_{T}^{\gamma} = p_{T}^{parton}$, 
modulo negligible corrections from initial state radiation.
In addition, due to the large mean free path of 
the photon compared to the system size formed in heavy-ion collision, direct photons decouple from the 
medium upon creation without any further interaction with the medium, and therefore the direct-photon 
measurement doesn't suffer from the same geometrical bias of that of single particle spectra and di-hadron 
azimuthal correlation measurements. In particular the direct $\gamma$-hadron azimuthal correlations provide a unique 
way to quantify the energy-loss dependence on the initial parton energy and possibly the color factor [12]. A comparison of direct
$\gamma$-hadron azimuthal correlations with di-hadron azimuthal correlations impart better constrain on the path length dependence
of energy loss; since the former average over the volume while the latter average over the surface of the formed medium. 

\section{Data Analysis}

This paper reports on the measurements of azimuthal correlations at mid-rapidity of direct photons at 
high transverse momentum (8 $< p_T^{\gamma, \pi^{0}} <$ 16 $\mathrm GeV$) with away-side charged hadrons of 
transverse momentum (3 $< p_{T}^{h^{\pm}} <$ 6 $\mathrm GeV/c$) over a broad range of centralities for 
$Au+Au$ collisions and $p+p$ collisions at $\sqrt{s_{NN}}$ = 200 $\mathrm GeV$ in the STAR experiment. 
Unfortunately direct photons have a rather small cross-section, making it a difficult measurement requiring a 
high-statistics data set. Using a level-2 high-$E_T$ tower trigger to tag $\gamma$-jet events, in 2007 the 
STAR experiment collected an integrated luminosity of 535 $\mu {b}^{-1}$ of $Au+Au$ collisions at 
$\sqrt{s_{NN}}$ = 200 $\mathrm GeV$. The level-2 trigger algorithm was implemented in the Barrel Electromagnetic
Calorimeter (BEMC) [13] and optimized based on the information of the direct $\gamma/\pi^0$ ratio in $Au+Au$ 
collisions [14], the $\pi^0$ decay kinematics, and the electromagnetic shower profile characteristics. The 
BEMC has full azimuthal coverage and pseudorapidity coverage of $\mid\eta\mid$ $\leq$ 1.0. As a reference 
measurement, we have analyzed $p+p$ data at $\sqrt{s_{NN}}$ = 200 $\mathrm GeV$ taken in 2006 with integrated luminosity 
of 11 $\mathrm pb^{-1}$. The Time Projection Chamber (TPC) [15] was used to detect charged-particle tracks and 
measure their momenta. The charged track quality cuts are similar to previous STAR analyses [16]. For this analysis, 
events with at least one cluster with $E_T >$ 8 $\mathrm GeV$ were selected. To ensure the purity of the 
photon-triggered sample, trigger towers were rejected for which a track with $p >$ 3 $\mathrm GeV/c$ pointed to it. To avoid the 
energy leakage near the edges of the BEMC acceptance in pseudorapidity, the trigger towers were restricted to a 
pseudorapidity of $\mid\eta\mid$ $\leq$ 0.9.

A crucial step of the analysis is to discriminate between showers of direct $\gamma$ and two close $\gamma$'s from a
high-$p_{T}$ $\pi^{0}$ symmetric decay. At $p_T$ $\sim$ 8 $\mathrm GeV/c$ the angular separation between the two photons 
resulting from a $\pi^{0}$ symmetric decay (both decay photons have similar energy, smallest opening angle) at the BEMC face is 
typically smaller than the tower size ($\Delta\eta=0.05,\Delta\phi=0.05$); but a $\pi^{0}$ shower is generally broader 
than a single $\gamma$ shower. The Barrel Shower Maximum Detector (BSMD) [13], which resides at approximately 5.6 
radiation lengths (X$_{0}$), at $\eta$ = 0, inside the calorimeter towers, is well-suited for 
$(2\gamma$)/$(1\gamma)$ separation up to $p_T$ $\sim$ 26 $\mathrm GeV/c$ due to its fine segmentation 
($\Delta\eta\approx 0.007,\Delta\phi\approx 0.007$). In this analysis the $\pi^{0}$/$\gamma$ discrimination was 
carried out by making cuts on the shower shape as measured by the BSMD, where the $\pi^{0}$ identification 
cut was adjusted in order to obtain a very pure sample of $\pi^{0}$ and a sample rich in direct $\gamma$ ($\gamma_{rich}$). 
The discrimination cuts were varied to determine the systematic uncertainties. 

To determine the combinatorial background level the relative azimuthal angular distribution of the associated 
particles with respect to the trigger particle is fitted with two Gaussian peaks and a straight line.  
The near- and away-side yields, $Y^{n}$ and $Y^{a}$, of associated particles per trigger are extracted by 
integrating the $\mathrm 1/N_{trig} dN/d(\Delta\phi)$ distributions in $\mid\Delta\phi\mid$ $\leq$  0.63 
and $\mid\Delta\phi -\pi\mid$  $\leq$  0.63 respectively. The yield is corrected for the tracking efficiency 
of associated charged particles as a function of multiplicity.

The shower shape cuts used to select a sample of direct-photon-``rich" triggers reject most of the $\pi^{0}$'s, 
but do not reject photons from highly asymmetric $\pi^{0}$ decays, $\eta$'s, and fragmentation photons. 
All of these sources of background are removed as follows (Eq. 2.1), but only within the systematic
uncertainty resulting from the assumption that their correlations are similar to those for $\pi^{0}$'s.
Assuming zero near-side yield for direct photon triggers and a very pure sample of $\pi^{0}$, which is verified 
by the $\pi^{0}$-charged yields in Fig. 3, the away-side yield of hadrons correlated with the direct photon is 
extracted as
\begin{equation}%\begin{eqnarray}
Y_{\gamma_{direct}+h}=\frac{Y^{a}_{\gamma_{rich}+h}-{\cal{R}} Y^{n}_{\gamma_{rich}+h}}{1-R},\label{eq:one}
\hspace{0.3cm}                            
   \hspace{0.3cm}   {\cal{R}}=\frac{Y^{a}_{\pi^{0}+h}}{Y^{n}_{\pi^{0}+h}};\hspace{0.3cm} and \hspace{0.3cm} 
   1-R={\frac{{N^{\gamma_{dir}}}}{N^{\gamma_{rich}}}}.
\end{equation}
Here $Y^{a(n)}_{\gamma_{rich}+h}$ and $Y^{a(n)}_{\pi^{0}+h}$ are the away (near)-side yields of associated particles 
per $\gamma_{rich}$ and $\pi^{0}$ triggers, respectively, the ratio $R$ is equivalent to the 
fraction of ``background'' triggers ($\pi^{0}$, $\eta$, and fragmentation photons) in the $\gamma_{rich}$ trigger sample, and 
$N^{\gamma_{dir}}$ and $N^{\gamma_{rich}}$ are the trigger number of direct $\gamma$ and $\gamma_{rich}$ respectively.

\section{Results}

\begin{figure}
\begin{tabular}{cc}
   \resizebox{85mm}{!}{\includegraphics{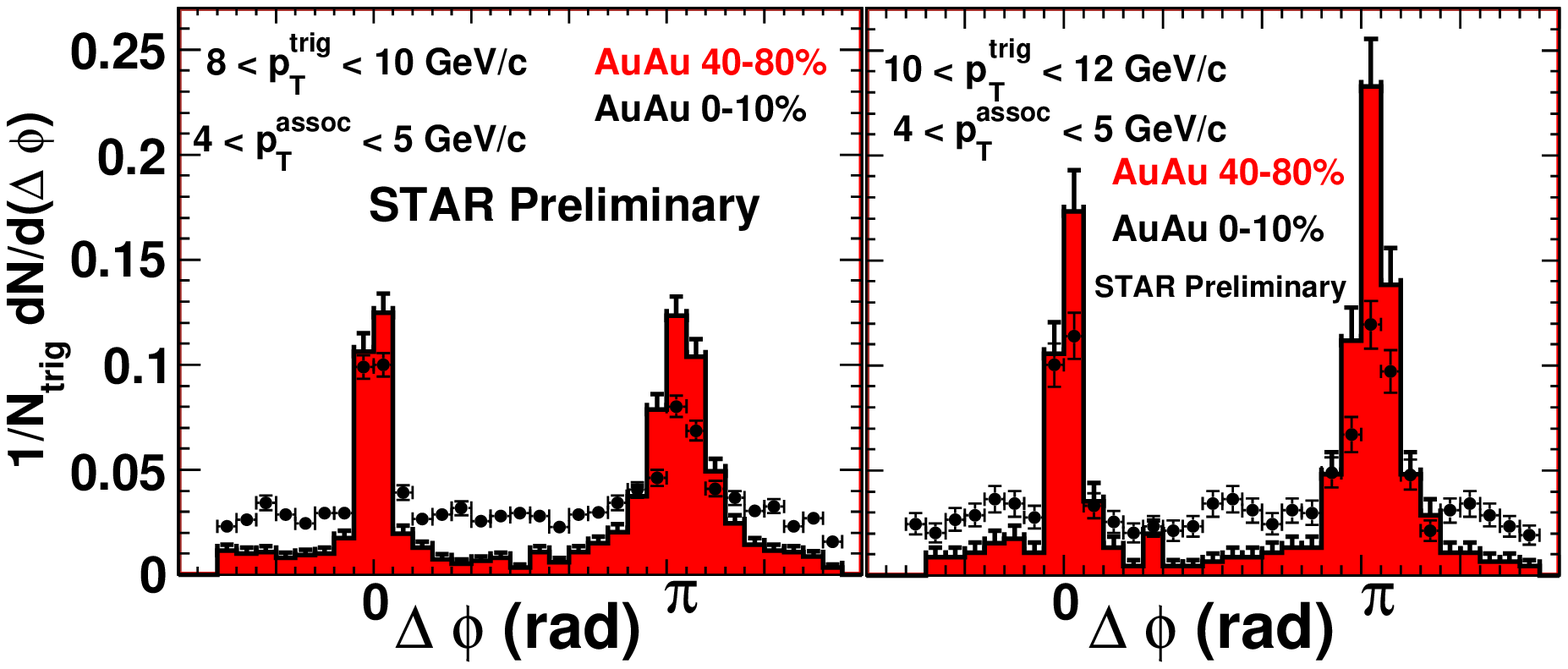}} 
    \resizebox{65mm}{!}{\includegraphics{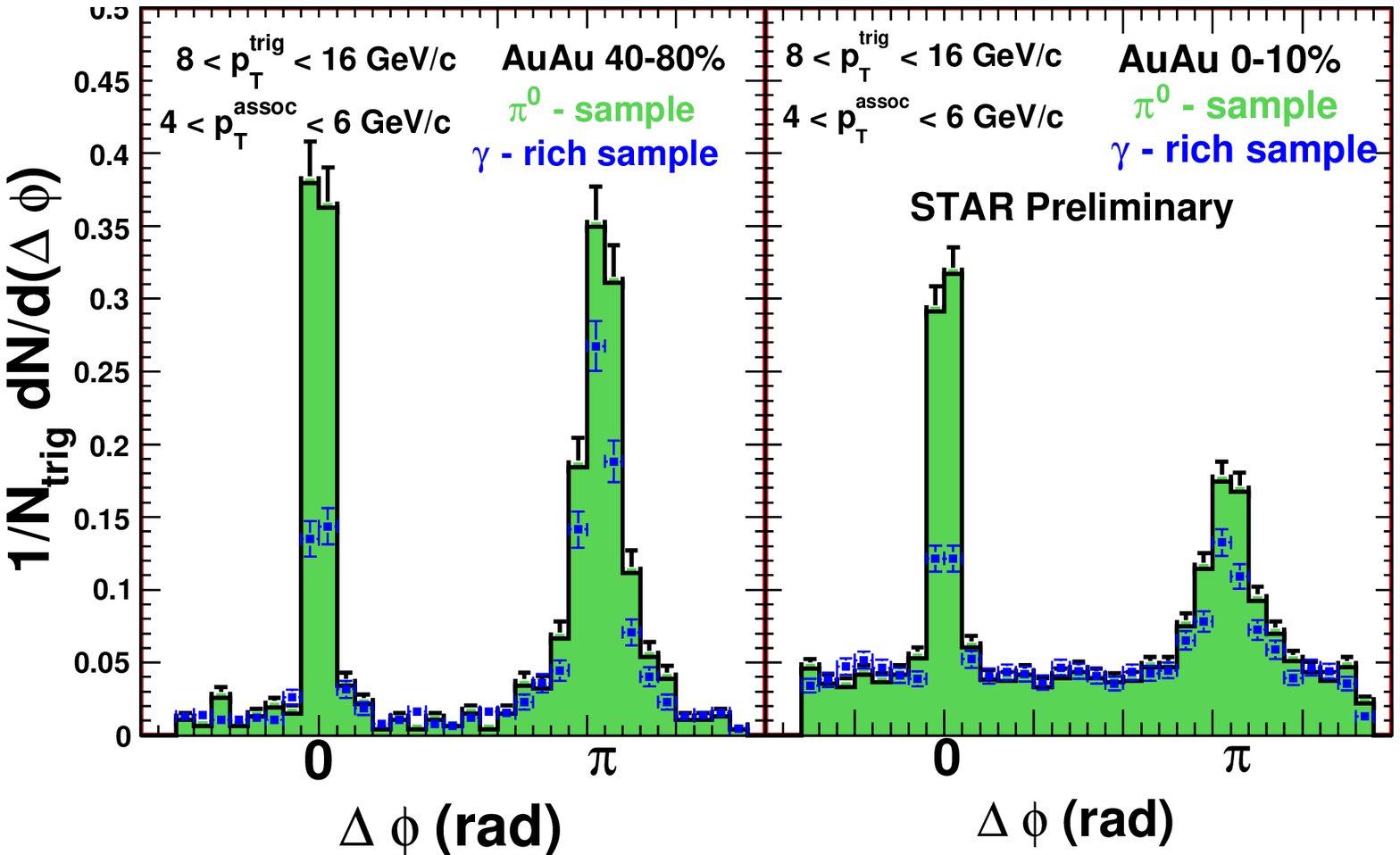}} \\
        \end{tabular}
    \caption{Left: Azimuthal correlation histograms of high $p_{T}^{trig}$ inclusive photons 
    with associated hadrons for 40-80$\%$ and 0-10$\%$ $Au+Au$
    collisions at $\sqrt{s_{NN}}$ = 200 $\mathrm GeV$. Right: Azimuthal correlation histograms of high $p_{T}^{trig}$ $\gamma_{rich}$ sample and $\pi^{0}$-sample  
    with associated hadrons for 40-80$\%$ and 0-10$\%$ $Au+Au$
    collisions at $\sqrt{s_{NN}}$ = 200 $\mathrm GeV$}
      \label{fig:1} 
\end{figure}
Figure 1 (left) shows the azimuthal correlation for inclusive photon triggers for the most peripheral and central bins in $Au+Au$ collisions. 
Parton energy loss in the medium causes the away-side to be increasingly suppressed with centrality as it was previously reported
[16,17].
The suppression of the near-side yield with centrality, which has not been observed in the charged hadron 
azimuthal correlation, is consistent with an increase of the $\gamma$/$\pi^{0}$ ratio with centrality at high $E_{T}^{trig}$. 
The shower shape analysis is used to distinguish between the $(2\gamma$)/$(1\gamma)$ showers as in Figure 1 (right) 
which shows the azimuthal correlation for $\gamma_{rich}$ sample triggers and $\pi^{0}$ triggers for the
most peripheral and central bins. The $\gamma_{rich}$ sample has lower near-side yield than $\pi^{0}$ but not zero.
The non-zero near-side yield for the $\gamma_{rich}$ sample is expected due to the remaining contributions of the widely 
separated photons from other
sources. The shower shape analysis is only effective for the two close $\gamma$ showers. 

The purity of using the shower shape analysis in $\pi^{0}$
identification is verified by comparing to previous measurements of azimuthal correlations between charged hadrons ($ch-ch$) [16]. 
\begin{figure}
\begin{center}
\begin{tabular}{cc}
   \resizebox{120mm}{!}{\includegraphics{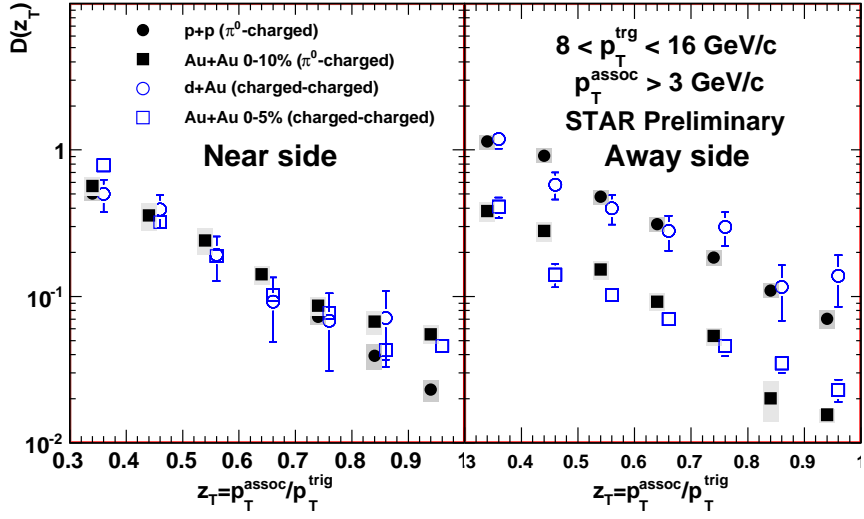}} 
        \end{tabular}
    \caption{$z_{T}$ dependence of $\pi^{0}-ch$ and $ch-ch$ [16] near-side (left panel) and away-side (right panel) associated particle yields.
    All collisions are at $\sqrt{s_{NN}}$ = 200 $\mathrm GeV$.}
      \end{center} 
      \label{fig:2} 
\end{figure}
Figure 2 shows the $z_{T}$ dependence of the associated hadron yield normalized per $\pi^{0}$ trigger $D(z_{T}$), where 
$z_{T}= p_{T}^{assoc}/p_{T}^{trig}$
[18], for the near-side and away-side compared to
the per charged hadron trigger [16]. The near-side yield as in
Figure 2 (left) shows no significant difference over the shown $z_{T}$ range between $p+p$, $d+Au$, and $Au+Au$ indicating in-vacuum fragmentation even in 
heavy ion collisions, which reveals
the surface bias as generated in several model calculations [19-22].
However the medium effect is clearly 
seen in the away-side in Figure 2 (right) where the per trigger yield in
$Au+Au$ is significantly suppressed compared to $p+p$ and $d+Au$. The general agreement between 
the results from this analysis
($\pi^{0}-ch$) and the previous analysis ($ch-ch$) is clearly seen in both panels of Figure 2 
which indicates the
purity of the $\pi^{0}$ sample and therefore the effectiveness of shower shape analysis in $\pi^{0}$ identifications.
An overall agreement for the $p_{T}^{assoc}$
dependence of the near-side associated yields for ($\pi^{0}-ch$) correlations compared to that of  
($ch-ch$) have been reported earlier [23]. 
\begin{figure}
\begin{center}
\begin{tabular}{cc}
   \resizebox{120mm}{!}{\includegraphics{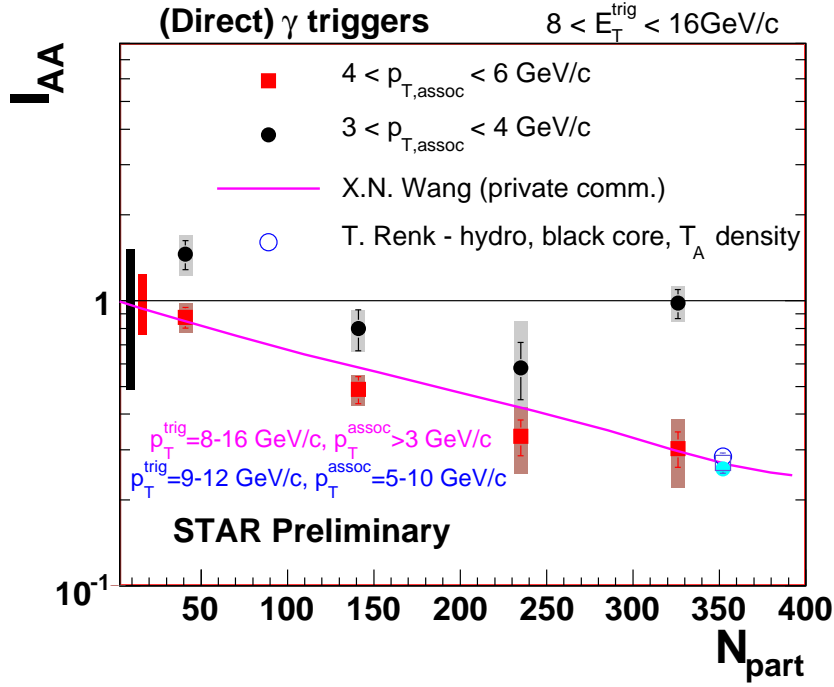}} 
        \end{tabular}
    \caption{$I_{AA}$ for direct $\gamma $ triggers (see text). Boxes on the left show the scale uncertainty due to $p+p$ measurements.
    All collisions are at $\sqrt{s_{NN}}$ = 200 $\mathrm GeV$}
      \end{center} 
      \label{fig:15} 
\end{figure}

The away-side yield of hadrons correlated with the direct $\gamma$ triggers (8 $<$ $p_T^{\gamma}$ $<$ 16 $\mathrm GeV$) is 
extracted as indicated by Eq. 2.1 in $p+p$ and $Au+Au$ collisions for (3 $<$ $p_T^{h^{\pm}}$ $<$ 4 $\mathrm GeV/c$) 
and (4 $<$ $p_T^{h^{\pm}}$ $<$ 6 $\mathrm GeV/c$) of charged hadrons. In order to quantify the away-side suppression, 
we calculate the quantity $I_{AA}$, which is defined as the ratio of the integrated yield of the away-side associated 
particles per trigger particle (direct $\gamma$) in $Au+Au$ relative to $p+p$ collisions. Figure 3 shows the 
$I_{AA}^{\gamma h^{\pm}}(N_{part.})$ for direct $\gamma$ triggers as a function of centrality.
The ratio would be unity if there were no medium effects on the parton fragmentation; and indeed the most
peripheral bin shows a ratio close to unity. The ratio decreases with centrality for more central events in a 
similar fashion for different ranges of $p_{T}^{assoc}$. The statistical and systematic uncertainties on 
the $p+p$ measurement result in a rather large scale uncertainty, which can be reduced with larger data samples in the future. 
The value of $I_{AA}^{\gamma h^{\pm}}$ for $\gamma$-hadron correlations in the most central
events is found to be similar to the values observed for di-hadron $I_{AA}^{h^{\pm} h^{\pm}}$ correlations and for 
single-particle suppression ($R_{AA}^{h}$) [24,25], and agrees well with theoretical calculations in which the energy 
loss is tuned to the single- and di-hadron measurements [26,27]. 

Figure 4 (left) shows the $z_{T}$ dependence of the trigger-normalized fragmentation function for $\pi^{0}-$charged 
correlations ($\pi^{0}-ch$) compared to measurements with direct $\gamma-$charged correlations ($\gamma-ch$). 
The away-side yield per trigger of direct $\gamma$ is smaller than with $\pi^{0}$ trigger at the same centrality class. 
This difference can be due to the fact that the $\pi^{0}$ originates from higher initial parton energy and therefore supports the 
energy loss at the parton level before fragmentation as it was previously reported [28] or it can be explained if 
a large fraction of direct $\gamma$ is produced via $qg \rightarrow \gamma q$ constituent scattering.  
\begin{figure}
 \begin{center}
  \begin{tabular}{cc}
   \resizebox{75mm}{!}{\includegraphics{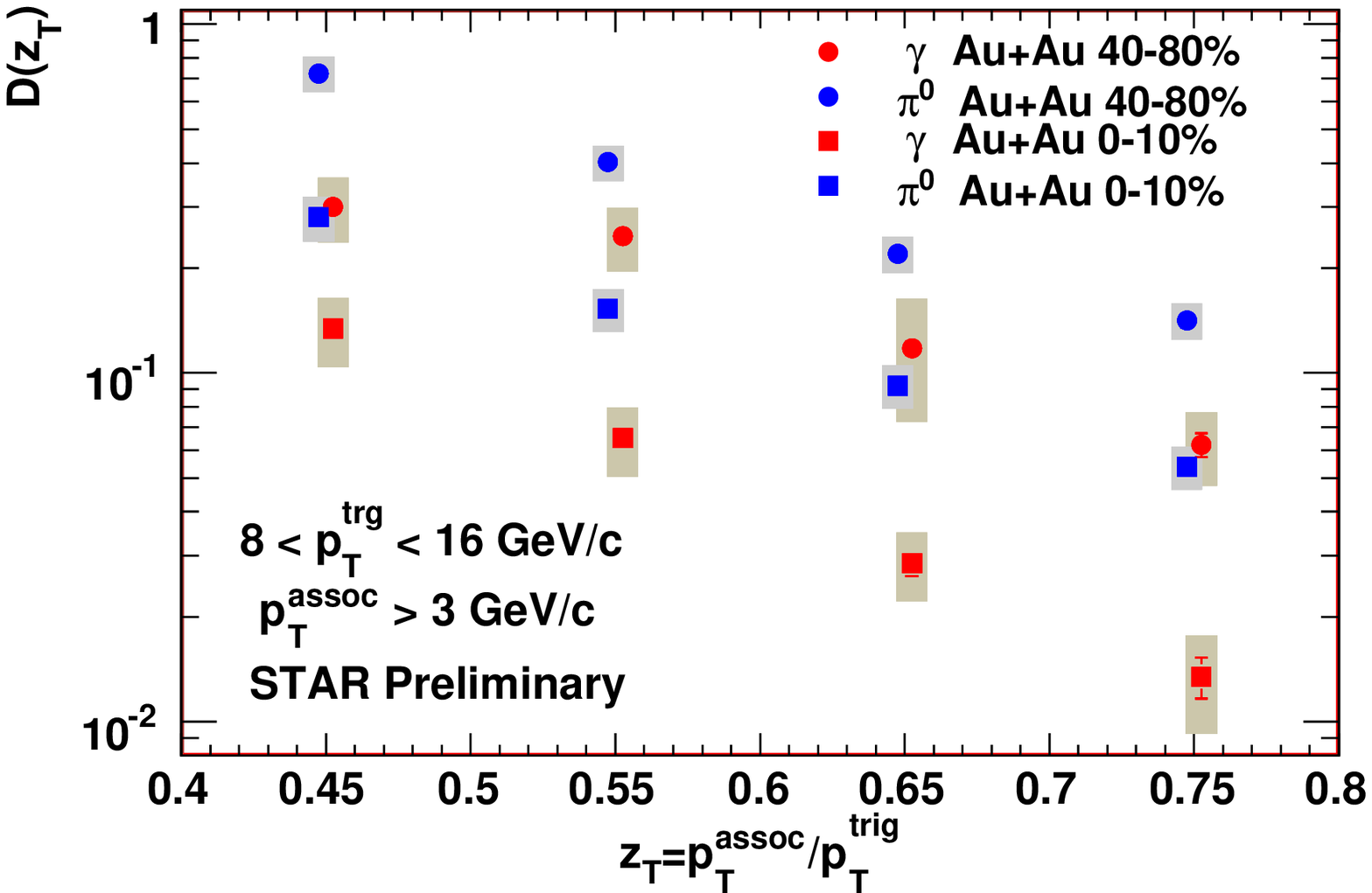}} 
   \resizebox{75mm}{!}{\includegraphics{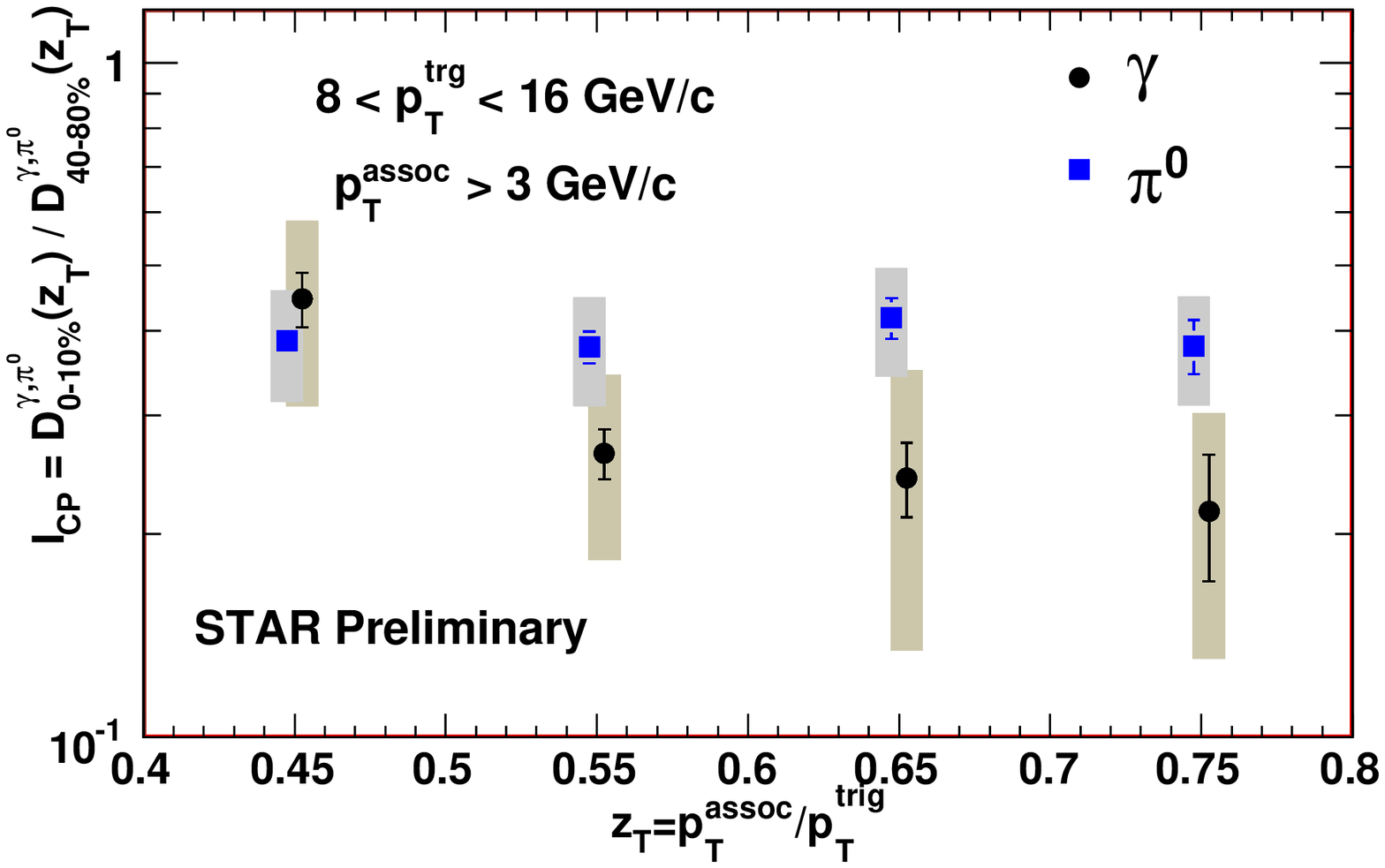}} \\
     \end{tabular}
    \caption{(Left) $z_{T}$ dependence of $\pi^{0}$ triggers and direct $\gamma$ triggers associated particle yields for 40-80$\%$ and
    0-10$\%$ $Au+Au$ collisions at $\sqrt{s_{NN}}$ = 200 $\mathrm GeV$. (Right) $z_{T}$ dependence of $I_{CP}$ for direct $\gamma$ triggers and $\pi^{0}$ triggers (see text). Boxes
    show the systematic uncertainties.}
    \end{center}
    \label{fig:3}
\end{figure}
\begin{figure}
 \begin{center}
  \begin{tabular}{cc}
   \resizebox{75mm}{!}{\includegraphics{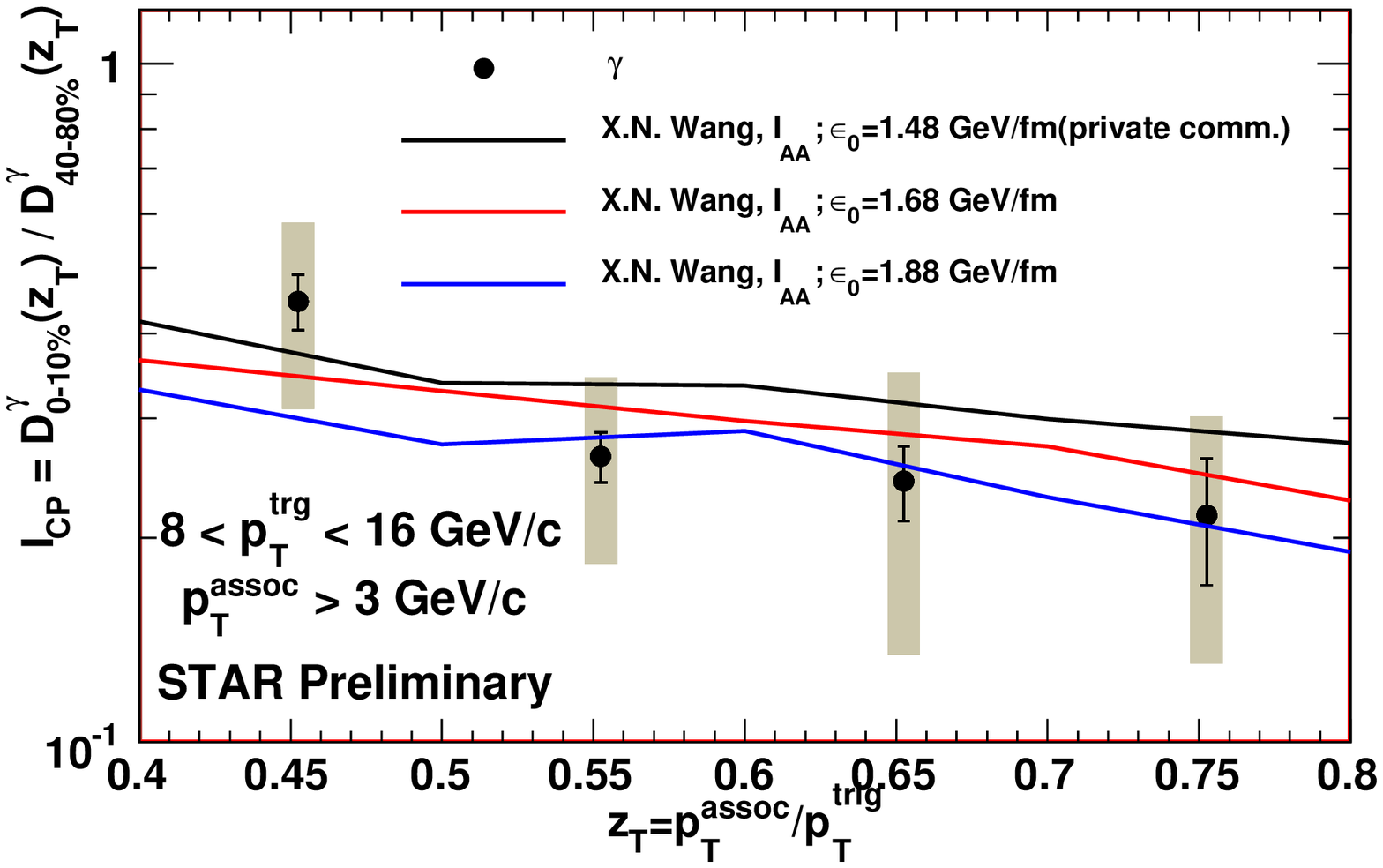}} 
   \resizebox{75mm}{!}{\includegraphics{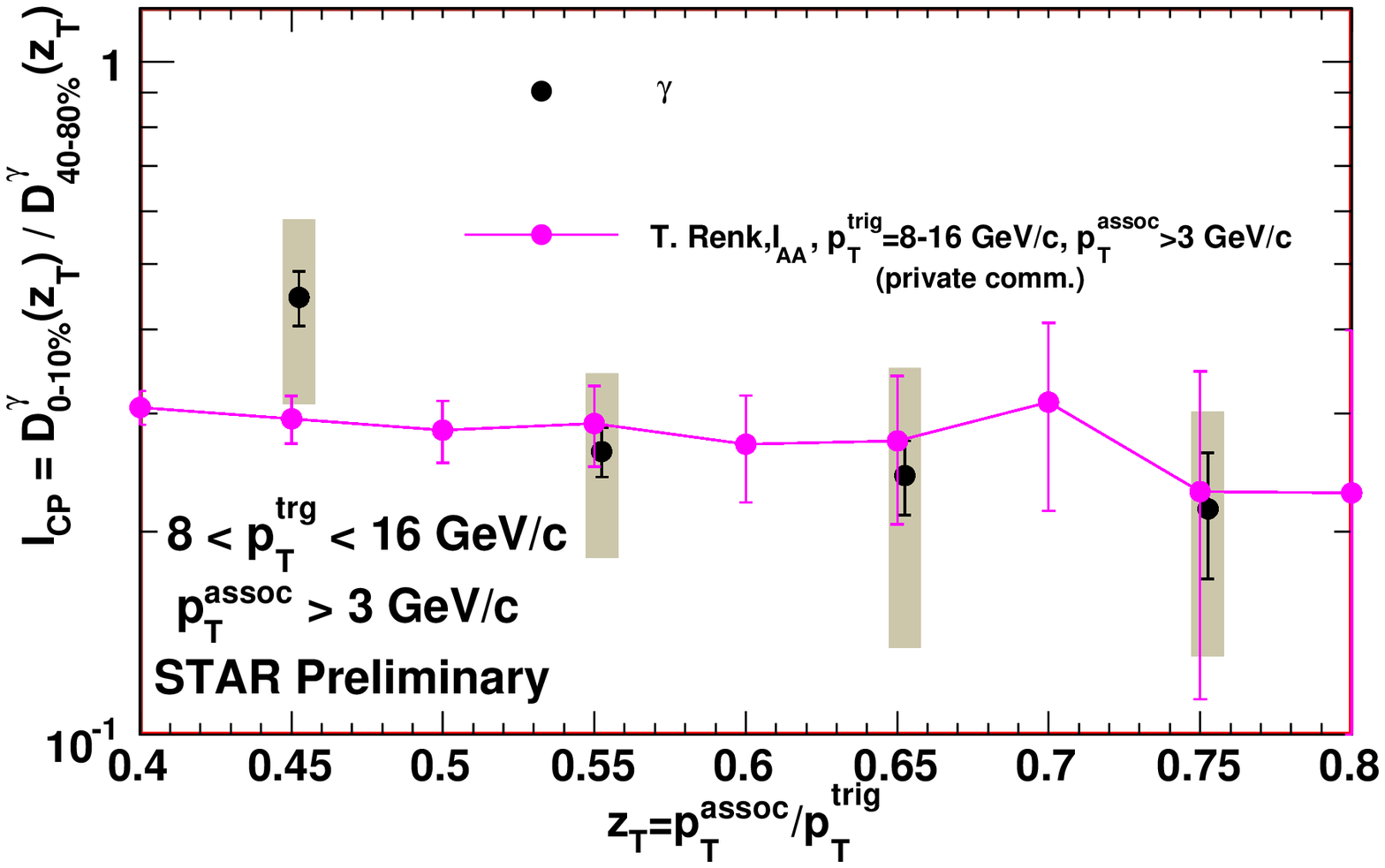}} \\
     \end{tabular}
    \caption{$z_{T}$ dependence of $I_{CP}$ for direct $\gamma$ triggers associated particle yields compared with theoretical calculations
    (left) $I_{AA}$ of 0-10$\%$ $Au+Au$ collisions (Only Annihilation and Compton processes to NLO are 
    considered) at $\sqrt{s_{NN}}$ = 200 $\mathrm GeV$ with three different initial gluon density where 7 $<$ $p_T^{trig}$ $<$ 9 $\mathrm GeV/c$ and $p_T^{assoc}$ $>$ 5 $\mathrm
    GeV/c$, (Right) $I_{AA}$ where 
    8 $<$ $p_T^{trig}$ $<$ 16 $\mathrm GeV/c$ and $p_T^{assoc}$ $>$ 3 $\mathrm GeV/c$.}
    \end{center}
    \label{fig:4}
\end{figure}

In order to quantify the away-side suppression, we calculate the quantity $I_{CP}$, which is defined as the 
ratio of the integrated yield of the away-side associated particles per trigger particle in $Au+Au$ central 
0-10$\%$ of the geometrical cross section; relative to $Au+Au$ peripheral 40-80$\%$ of the geometrical cross 
section collisions. Figure 4 (right) shows the $I_{CP}$ for $\pi^{0}$ triggers and for direct $\gamma$ triggers 
as a function of $z_{T}$. The medium effect on the parton fragmentation is obvious; 
and indeed the ratio deviates from unity by a factor of $\sim$ 2.5. The ratio for the $\pi^{0}$ trigger is 
approximately independent of $z_{T}$ for the shown range in agreement with the previous results from ($ch-ch$) measurements [16]. 
Within the current systematic uncertainty the $I_{CP}$ of direct $\gamma$ and $\pi^{0}$ are similar.    

Suppression ratios with respect to the p+p reference, $I_{AA}$, have
been shown in Fig. 3. The values of $I_{AA}$ are smaller than for $I_{CP}$,
indicating finite suppression in the peripheral 40-80$\%$ data, but the
statistical uncertainties are large due to the small $\gamma$/$\pi^{0}$ ratio in $p+p$ as previously reported [29]. Although 
the value of $I_{AA}$ is found to be similar to the values observed
for di-hadron correlations and for single-particle suppression $R_{AA}$.  

A comparison of $I_{CP}$ of direct $\gamma$ with two different theoretical model calculations of $I_{AA}$ of 
direct $\gamma$ is shown in Figure 5.
The $I_{CP}$ values agree well with the two different theoretical predictions within the current uncertainties.
Although the two models work in the pQCD framework but the detailed calculations and the medium modeling are quite different. 
Figure 5 (left) indicates the need for more reduction in the systematic and statistical uncertainties in order 
to distinguish between different color charge densities.   

\section{Summary and Outlook}
In summary, the 
$\gamma-$charged hadron coincidence measurements in $Au+Au$ collisions have been performed and reported by the 
STAR experiment. The STAR detector is unique to perform such correlation measurements 
due to the full coverage in azimuth. 
The $I_{AA} ^ {\gamma h^{\pm}}$ and $I_{AA} ^ {{\pi^{0}} h^{\pm}}$ 
show similar suppression to that of 
$I_{AA} ^ {h^{\pm} h^{\pm}}$, $R_{AA} ^ {\pi^{0}}$, and $R_{AA} ^ {h^{\pm}}$ of light and heavy quarks.
The energy loss dependence of parton initial energy can be studied precisely through the 
$\gamma-$charged hadron results.
Within the current uncertainty the recoil suppression ratio $I_{CP}$ of direct $\gamma$ and $\pi^{0}$ trigger are similar, which doesn't
allow to study the path length dependence of energy loss. This can be better studied via $\gamma-$hadron coincidence measurements in- and
out- of reaction plane. 
A full analysis of the systematic uncertainties is under way and may
lead to a reduction of the total uncertainty. Future RHIC runs will
provide larger data samples to further reduce the uncertainties and
extend the $z_{T}$ range.

\section{Acknowledgments}

I would like to thank the organizers for support and interesting discussions.


\begin{thebibliography}{99}
\bibitem{1} X. N. Wang and M. Gyulassy, Phys. Rev. Lett. 68, 1480 (1992).
\bibitem{2} R. Baier et al., Z. Physik C, Particles and Fields 6, 309-316 (1980) and references therein.
\bibitem{3} S. Adler et al., Phys. Rev. Lett.91 072303 (2003).
\bibitem{4} Aidala C et al. http://spin.riken.bnl.gov/rsc/report/masterspin.pdf (Research Plan for Spin Physics
at RHIC); update: Bunce G et al. http://www.physics.rutgers.edu/np/RHIC spin LRP07.pdf.
\bibitem{5} J. F. Owens, Rev. Mod. Phys. 59, 465 (1987). 
\bibitem{6} T. Ferbel and W. R. Molzon, Rev. Mod. Phys. 56, 181 (1984).
\bibitem{7} P. Aurenche and J. Lindfors, Nucl. Phys. B 168, 296 (1980).
\bibitem{8} M. Fontannaz and D. Schiff, Nucl. Phys. B 132, 457 (1978).
\bibitem{9} A. P. Contogouris et al., Phys. Rev. D 32, 1134 (1985).
\bibitem{10} M. D. Corcoran et al., Phys. Rev. Lett. 41, 9 (1978); M. Dris et al., Phys. Rev. D 19, 1361 (1979).
\bibitem{11} J. F. Owens, Phys. Rev. D 20, 221 (1979).
\bibitem{12} B. Mohanty (STAR Collaboration) J. Phys. G: Nucl. Part. Phys. 35 104006 (2008), arXiv: 0804.4760v1. 
\bibitem{13} M. Beddo et al. , Nucl. Instrum. Meth. A 499 725 (2003).
\bibitem{14} K. Filimonov, Acta Phys.Hung. A25:363-370 (2006).
\bibitem{15} M. Anderson et al. , Nucl. Instrum. Meth. A 499 659 (2003).
\bibitem{16} J. Adams et al., Phys. Rev. Lett. 97 162301 (2006).
\bibitem{17} J. Adams et al., Phys. Rev. Lett.91 072304 (2003).
\bibitem{18} X.N.Wang, Phys. Lett. B 595, 165 (2004).
\bibitem{19} A. Drees, H. Feng and J. Jia, Phys. Rev. C 71, 034909 (2005).
\bibitem{20} A. Dainese, C. Loizides and G. Paic, Eur. Phys. J. C38, 461 (2005).
\bibitem{21} K. J. Eskola et al., Nucl. Phys. A 747, 511 (2005).
\bibitem{22} B. Muller, Phys. Rev. C 67, 061901 (2003).
\bibitem{23} A. M . Hamed (STAR Collaboration) J. Phys. G: Nucl. Part. Phys. 35 104120 (2008), arXiv: 0806.2190. 
\bibitem{24} B. I. Abelev et al., Phys. Lett. B 655, 104-113 (2007).
\bibitem{25} B. I. Abelev et al., Phys. Rev. Lett.97 152301 (2006).
\bibitem{26} X. N. Wang and H. Zhang, private communication. 
\bibitem{27} T. Renk, Phys. Rev. C74 034906 (2006). 
\bibitem{28} O. Catu (STAR Collaboration), J. Phys. G: Nucl. Part. Phys. 35 104088 (2008). 
\bibitem{29} S. Chattopadhyay (STAR Collaboration) J. Phys. G: Nucl. Part. Phys. 34, S985-S988 (2007). 
\end{thebibliography}
\end{document}